# Atlas of the Light Curves and Phase Plane Portraits of Selected Long-Period Variables

Kudashkina L.S., Andronov I.L.

Department of Mathematics, Physics and Astronomy, Odessa National Maritime University
*kuda2003@ukr.net, tt_ari@ukr.net*

For a group of the Mira-type stars, semi-regular variables and some RV Tau-type stars the limit cycles were computed and plotted using the phase plane diagrams. As generalized coordinates $x$ and $\dot{x}$, we have used $m$ – the brightness of the star and its phase derivative.

We have used mean phase light curves using observations of various authors from the databases of AAVSO, AFOEV, VSOLJ, ASAS and approximated using a trigonometric polynomial of statistically optimal degree. As generalized coordinates $x$ and $\dot{x}$, we have used $m$ – the brightness of the star and its phase derivative.

For a simple sine-like light curve, the limit cycle is a simple ellipse. In a case of more complicated light curve, in which harmonics are statistically significant, the limit cycle has deviations from the ellipse.

In an addition to a classical analysis, we use the error estimates of the smoothing function and its derivative to constrain an "error corridor" in the phase plane.

The pulsations of long-period variable stars – Mira-type, semi-regular variables RVab and one SRb star – are studied. This work is the continuation of a cycle of the works on the study of the photometric properties of the mean light curves of the long-period variables. The studied stars are located on the HR-diagram in the AGB (Mira-type and SR) and RGB (RVab) areas.

Earlier, for example, in the work (Kudashkina & Marsakova, 2013) for Mira-type stars R Aql, R Hya и T UMi, the time of the change of pulse period has been estimated "to zero" at the time of compression of the layers of the star due to the termination of helium burning in the layered source. Also the radius was estimated, from which begins the compression, for a range of stellar masses 0.9-1.5 $M_\odot$.

On the asymptotic giant branch (AGB), the star consists of a degenerate carbon-oxygen core and two layer sources (helium and hydrogen), positioned very close to each other. Above them is an extended hydrogen envelope.

The small thickness of the layered sources cause thermal flashes. Here, the stars are divided into two stages: early (EAGB) is the time interval between the end of helium burning in the core and the first thermal pulse of the helium layer source, and "thermally pulsing regime of the He-burning shell" (TPAGB) is the thermally pulsed combustion mode in the helium layer. At the stage of a TPAGB the star is getting brighter in $M_{bol}$. Some theoretical calculations show that, for example, zirconium stars over a range of luminosity and temperature correspond to the early stages of AGB when helium burning occurs stationary. At the same time, several observational features say, rather, in favor of finding these stars in more advanced evolutionary stage on the AGB (Kudashkina, 2003).

The different studies show that long-period variables pulsate in a fundamental mode and in the first overtone. The latter is especially for semiregular variables, but a significant part of the Mira-type stars are the overtone pulsators (Fadeev, 1993).

Indirect dependence of the pulsation properties from surface luminosity is also seen in the various photometric dependencies. For example, in the article Kudashkina

(2015), the three-dimensional diagram of the photometric parameters: the period, the amplitude of light, the slope of the ascending branch of the light curve are obtained.

The most significant correlation shows the dependence of the slope of the ascending branch of the period and amplitude.

This study is the next step in the study of the photometric parameters of long-period variables – Mira-type stars and the semi-regular and relative objects – which can be used as an additional criterion the classification of these stars to the EAGB and TPAGB stages. The presented method is supplementary to that used for our studies and listed by Andronov et al. (2016).

It is known that the mode of oscillation corresponds to a periodic limit cycle, i.e. a closed phase trajectory to strive for all of the close trajectory over time. We calculated limit cycles for 8 long-period variables and 5 stars of RV Tau-type, based on the mean light curves, averaged over a long period of time. As generalized coordinates of the phase plane are taken $m$ – brightness of the star and its phase derivative. That is, the curve of evolution of the brightness of the star in the phase space is $m(\varphi)$.

For approximation of the phase curve, we have used trigonometric polynomials

$$m(\varphi) = C_1 + \sum_{j=1}^{M}(C_{2j}\cos(j\varphi) + C_{2j+1}\sin(j\varphi)), \qquad (1)$$

moreover, the optimal value of the degree $M$ is determined by the Fisher test with a critical probability of "false alarm" (FAP=False Alarm Probability) $10^{-3}$.

The phase derivative in radians calculated simple

$$\dot{m}(\varphi) = \sum_{j=1}^{M} j \cdot (-C_{2j}\sin(j\varphi) + C_{2j+1}\cos(j\varphi)), \qquad (2)$$

That is, upon differentiation, the amplitude of the harmonic is multiplied by $j$. Usually, this leads to a greater increase in the relative statistical error of the derivative compared to the signal.

Phase portraits of the stars made using an improved version of the program FDCN (Andronov, 1994) and are shown in Fig. 1-28 together with the approximations of their light curves.

Theoretically, for sinusoidal oscillations, the phase portrait is an ellipse with different units of measurement on axes – magnitude and magnitude "the radian" or "the period". Deviations from the ellipse expected for non-sinusoidal waves, i.e. if the degree of the trigonometric polynomial $M>1$. The calculation is performed in the cycle phase, the results for $m$, $\dot{m}$ are displayed with "3σ" corridors of errors on both coordinates. Naturally, the extreme points on each coordinate line "corridors of errors" for the other coordinate converge to a point like the letter Ж. Therefore, as the corridor of errors, use external part of the corridors of the errors for the two variables.

For the semiregular variable RV Tau limit cycle is almost a perfect ellipse, but for the value equals the half-period. For Mira-type stars, with more regular oscillations, but a complex form of the light curve, limit cycles deviate strongly from the ellipse. Other RV-type, Mira-type and semi-regular stars U Mon, X CrB, U UMi, X Oph, R Aql, U Her, S Ori, S Car, S Per, W Hya, $L_2$ Pup were investigated (Kudashkina & Andronov, 2017) and were also.

The list of stars and the basic data are given in table 1.

Table 1. The list of stars and its basic data

| N | Star | Type | Period, d |
|---|------|------|-----------|
| 1 | o Cet | M | 333.6 |
| 2 | S Scl | M | 367 |
| 3 | RR Aql | M | 390.78 |
| 4 | U Cyg | M | 465.49 |
| 5 | V Cyg | M | 421.27 |
| 6 | BG Cyg | M | 288.1 |
| 7 | AM Cyg | M | 371.9 |
| 8 | R Leo | M | 312 |
| 9 | Y CVn | SRb | 267 |
| 10 | RV Tau | RV | 78.73 |
| 11 | EP Lyr | RV | 82.95 |
| 12 | R Sge | RV | 71.57 |
| 13 | DF Cyg | RV | 776.4 |

For o Cet, S Scl and RR Aql, the periods have been determined using the Whitlock's H and K observations (Kudashkina, 2016). For Y CVn, the period has been updated by Kudashkina & Andronov (2010). For the rest of the star the values of the periods also have been updated (Kudashkina et al., 2013).

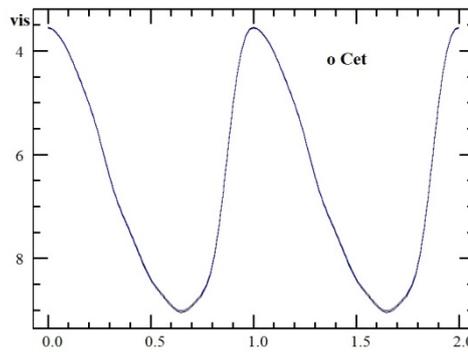 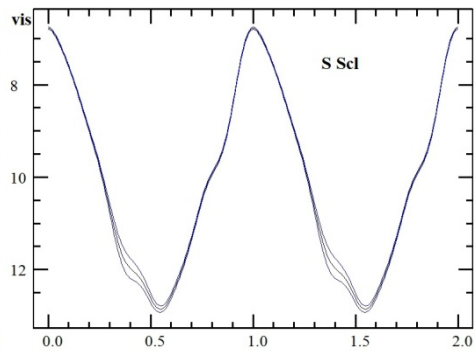

Fig. 1. The mean light curve of o Cet.    Fig. 2. The mean light curve of S Scl.

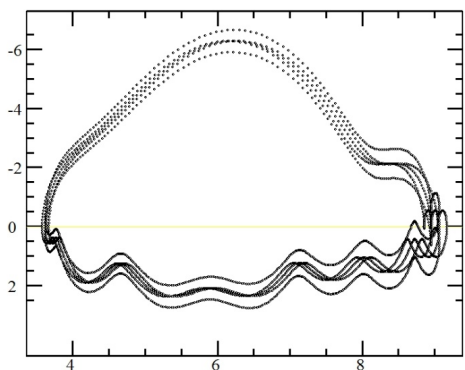 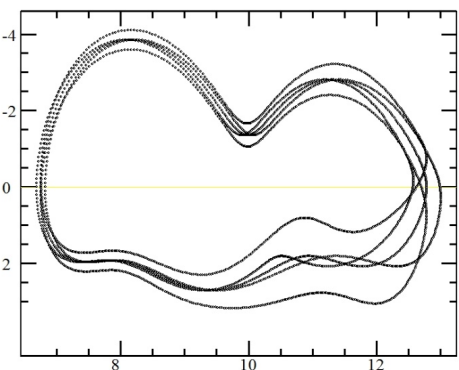

Fig. 3. Phase portrait of o Cet.    Fig. 4. Phase portrait of S Scl.

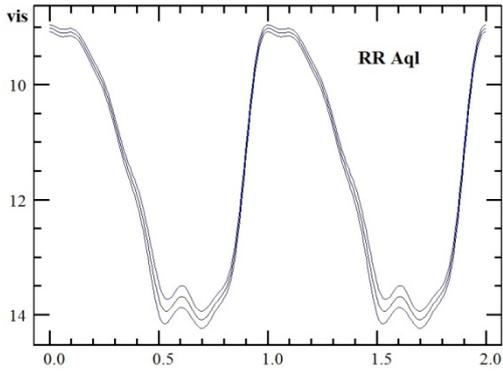
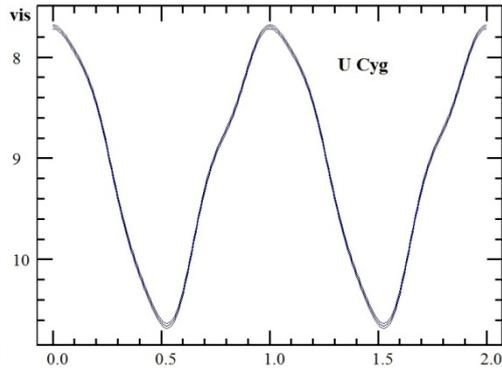

Fig. 5. The mean light curve of RR Aql.   Fig. 6. The mean light curve of U Cyg.

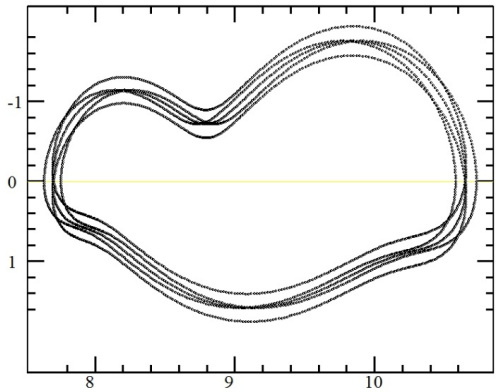
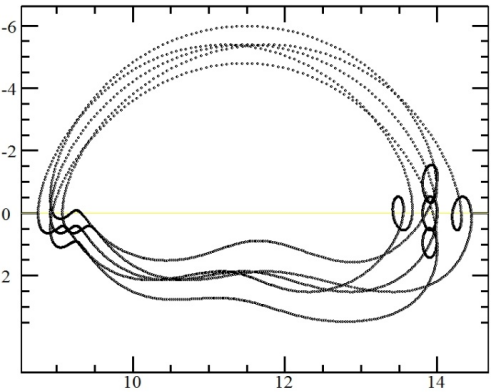

Fig. 7. Phase portrait of RR Aql.   Fig. 8. Phase portrait of U Cyg.

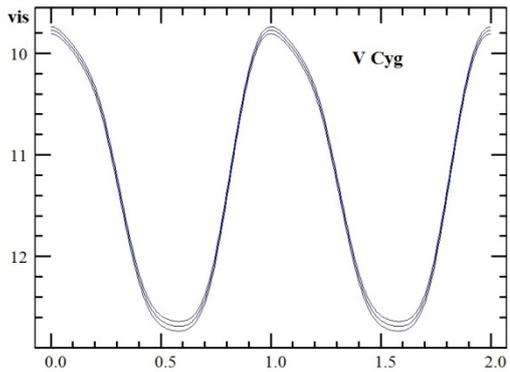
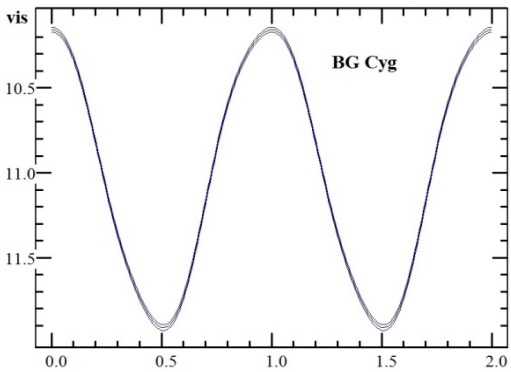

Fig. 9. The mean light curve of V Cyg.   Fig. 10. The mean light curve of BG Cyg.

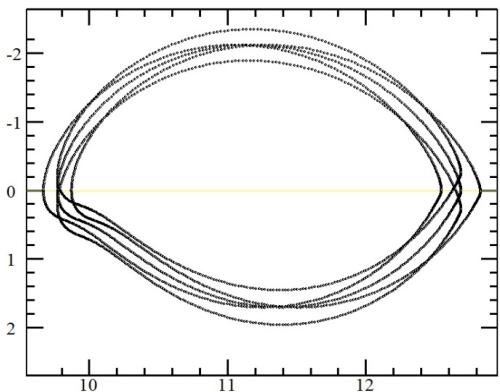
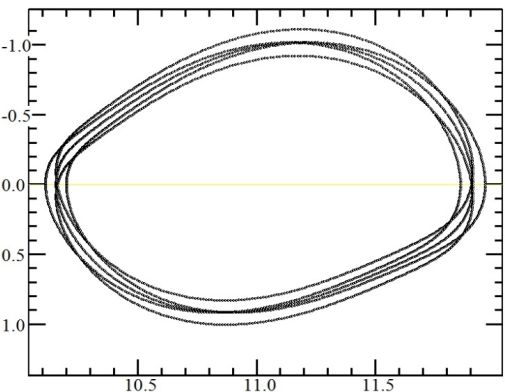

Fig. 11. Phase portrait of V Cyg.   Fig. 12. Phase portrait of BG Cyg.

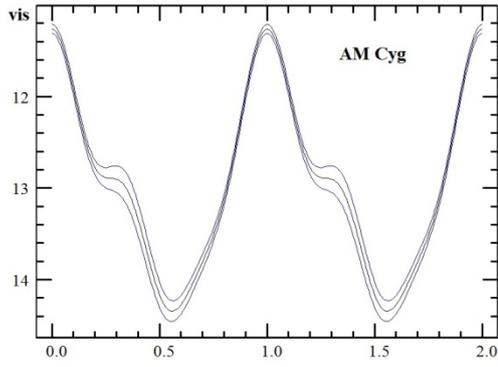
Fig. 13. The mean light curve of AM Cyg.

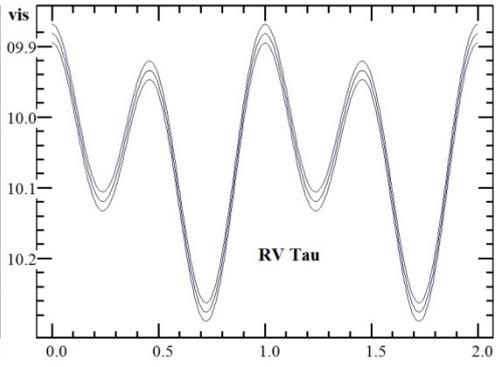
Fig. 14. The mean light curve of RV Tau.

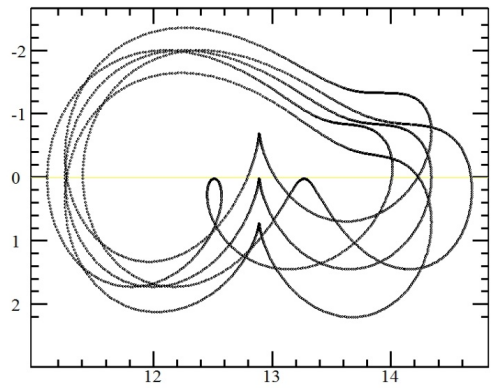
Fig. 15. Phase portrait of AM Cyg.

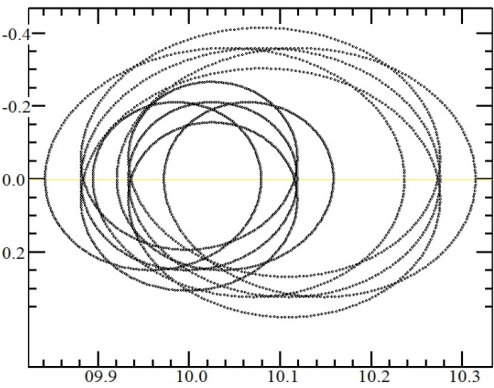
Fig. 16. Phase portrait of RV Tau.

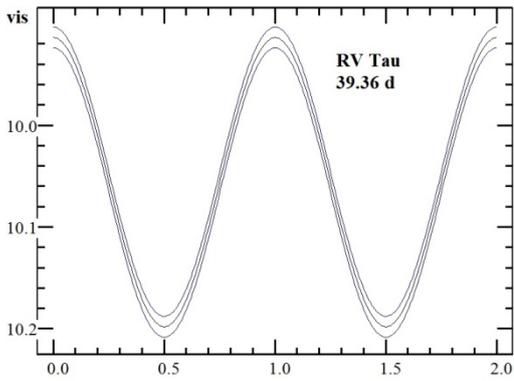
Fig. 17. The mean light curve of RV Tau with a half-period.

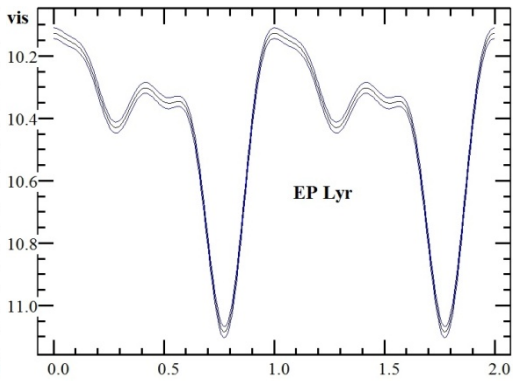
Fig. 18. The mean light curve of EP Lyr.

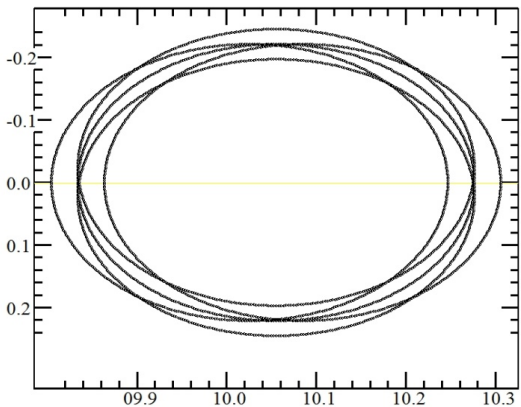
Fig. 19. Phase portrait of RV Tau with a half-period.

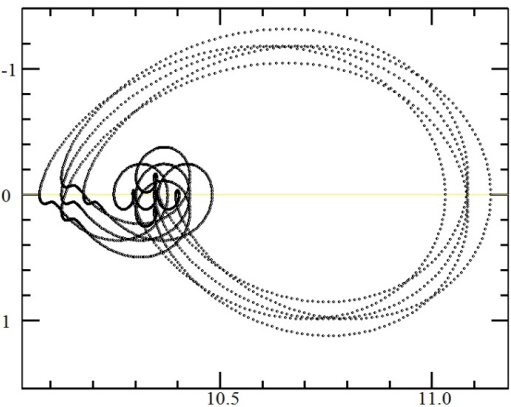
Fig. 20. Phase portrait of EP Lyr.

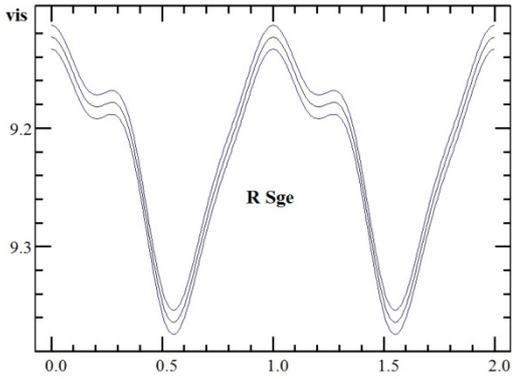
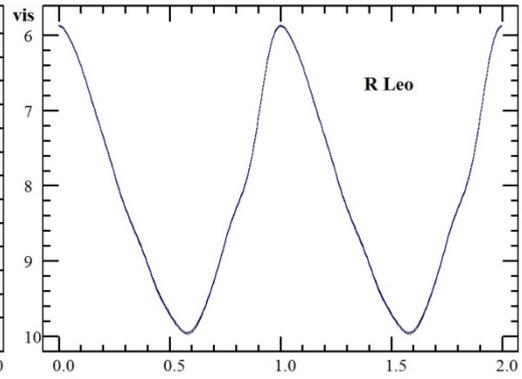

Fig. 21. The mean light curve of R Sge.   Fig. 22. The mean light curve of R Leo.

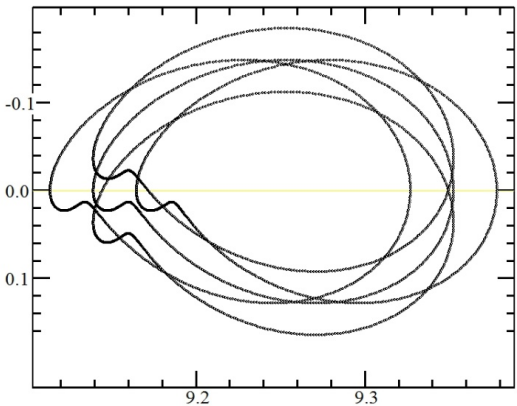
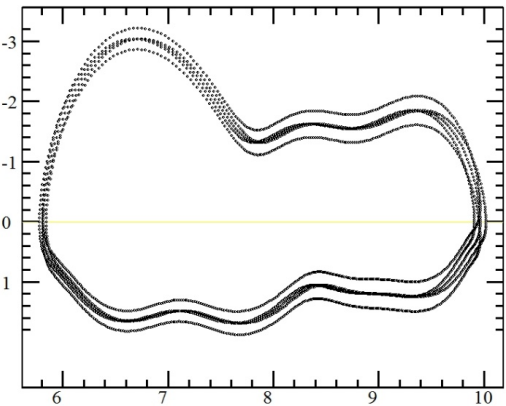

Fig. 23. Phase portrait of R Sge.   Fig. 24. Phase portrait of R Leo.

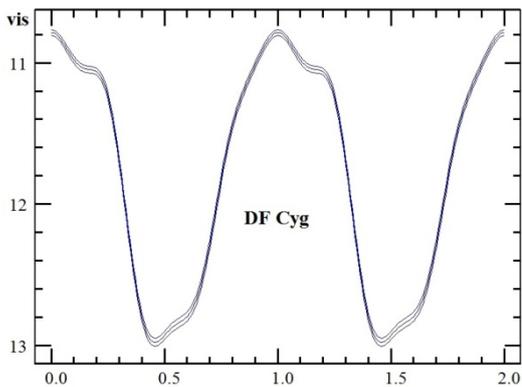
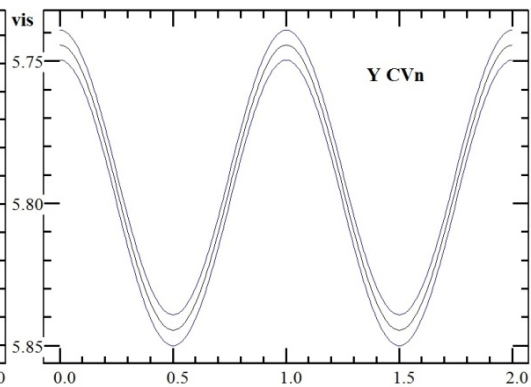

Fig. 25. The mean light curve of DF Cyg.   Fig. 26. The mean light curve of Y CVn.

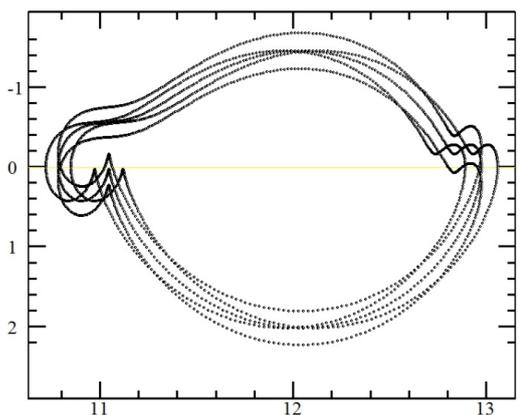
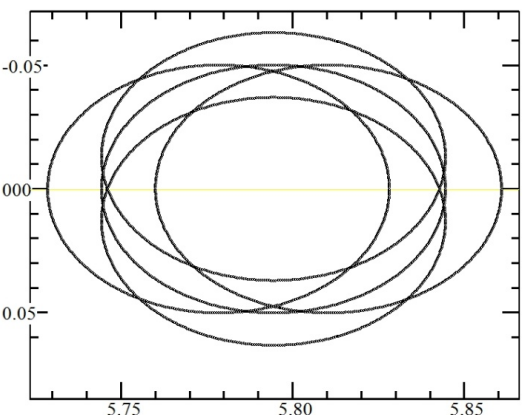

Fig. 27. Phase portrait of DF Cyg.   Fig. 28. Phase portrait of Y CVn.

One may note that the two-humped light curve of RV Tau phenomenologically resembles that of $\beta$ Lyrae. Generally, this method may be effectively applied to eclipsing binaries of various subtypes (see Tkachenko et al. 2016 for phenomenological modeling of prototype stars).

Further analysis is planned for a more extended sample of stars, including that included in the current "Atlas…".

These studies were carried out within the "Stellar Bell" part of the international campaign "Inter-Longitude Astronomy" (Andronov et al., 2003, 2014, 2017) and "Astroinformatics" (Vavilova et al., 2017).